\documentclass[conference,10pt]{IEEEtran}
\usepackage{xcolor}  % 用于颜色支持

\usepackage{graphicx}
\usepackage{subcaption}
\usepackage{float}
\usepackage{booktabs}
\usepackage{makecell}
\usepackage{amsmath}
\usepackage{xspace} 
\usepackage{amssymb}

\usepackage{pifont}% http://ctan.org/pkg/pifont
\author{
    \IEEEauthorblockN{Liming Liu, Jiangkai Wu, Xinggong Zhang}
    \IEEEauthorblockA{Peking University, China}
}

\title{Smaller is Better: Generative Models Can Power Short Video Preloading} 

\begin{document}

\maketitle
% \vspace{-5mm}

\begin{abstract}
% Short video platforms rely on preloading to ensure smooth playback, but face a persistent dilemma: aggressive preloading reduces stalls but wastes bandwidth, while conservative strategies save data but risk interruptions. We propose a novel preloading paradigm that uses local computation to reduce bandwidth requirements. By transmitting compact semantic prompts instead of pixel-level encodings, and decoding them with generative models like stable diffusion, we can reconstruct high-quality frames with far less data. We design a full-stack system with gradient-based prompt inversion, computation-aware scheduling, and a planning algorithm that scales to the enlarged decision space. Our approach reduces stalls and bandwidth waste by over 31\% and improves QoE by 45\%, demonstrating the promise of computation-powered preloading.

% \begin{abstract}
Preloading is widely used in short video platforms to minimize playback stalls by downloading future content in advance. However, existing strategies face a tradeoff. Aggressive preloading reduces stalls but wastes bandwidth, while conservative strategies save data but increase the risk of playback stalls. This paper presents \textbf{PromptPream}, a computation powered preloading paradigm that breaks this tradeoff by using local computation to reduce bandwidth demand. Instead of transmitting pixel level video chunks, PromptPream sends compact semantic prompts that are decoded into high quality frames using generative models such as Stable Diffusion. We propose three core techniques to enable this paradigm: (1) a \textit{gradient based prompt inversion} method that compresses frames into small sets of compact token embeddings; (2) a \textit{computation aware scheduling strategy} that jointly optimizes network and compute resource usage; and (3) a scalable \textit{searching algorithm} that addresses the enlarged scheduling space introduced by scheduler. Evaluations show that PromptStream reduces both stalls and bandwidth waste by over \textbf{31\%}, and improves Quality of Experience (QoE) by \textbf{45\%}, compared to traditional strategies. 
\end{abstract}

% \end{abstract}

% \pagestyle{plain}

\vspace{-3mm}
\section{Introduction}
\vspace{-2mm}

% background and dilemma
Short video platforms like TikTok and Instagram Reels are one of the most bandwidth-hungry applications in the mobile ecosystem~\cite{global-phenomena24}. To reduce perceived latency and improve user experience, these platforms rely heavily on \emph{preloading} chunks of videos before users actually watch them~\cite{phong2023joint,zhang2020apl}. Preloading can hide network stalls and smooth playback, but it also leads to a persistent and well-known dilemma: when bandwidth or user behavior prediction is inaccurate, the system either preloads too little and causes playback stalls~\cite{le2013buffer}, or preloads too much and wastes bandwidth on unseen content~\cite{zhang2022measurement}. This stall–waste tradeoff is not merely a tuning issue, it is fundamental to how existing preloading systems work: most systems make binary decisions on whether or not to preload a particular video chunk, with no mechanism to trade off chunk fidelity, size, or decode cost. 

% key insight
% In this paper, we explore a new paradigm: \emph{using local computation to reduce bandwidth requirements}, thereby mitigating both stalls and waste. Our key insight is that if video chunks can be made smaller \emph{without sacrificing perceptual quality}, then both stalls and bandwidth waste can be reduced, leading to significant improvements in user QoE. We demonstrate that this is possible by leveraging the viewer’s idle computational resources. Instead of transmitting verbose pixel-level encodings, we send compact semantic prompts that can be decoded by modern generative models like stable diffusion~\cite{Rombach_2022_CVPR} to reconstruct high quality frames.

In this paper, we explore a new paradigm: \emph{using local computation to reduce bandwidth requirements}, thereby mitigating both stalls and waste. Our key idea is to leverage the widely available GPU hardware, which remains almost idle during video decoding. While super-resolution\cite{yeo2018neural} or more advanced neural codecs\cite{jia2025towards} can also trade local computation for improved quality, we take this idea a step further. Instead of transmitting pixel-level encodings, we send compact \emph{semantic prompts} that can be decoded by modern generative models~\cite{Rombach_2022_CVPR} to reconstruct high-quality frames, offering greater potential for quality enhancement under the same bandwidth constraints.

% challenges
But turning this new idea into a practical system introduces several challenges. First, we need a method to convert video frames into minimal representations that preserve visual fidelity. Second, computation constraints invalidate traditional preloading strategies, requiring new mechanisms to evaluate and schedule chunks under both bandwidth, decode latency and computation resource limitations. Finally, we must make preload decisions across a large, interdependent space involving mixed codecs, varying decode times, and out of order execution, where the sheer number of options makes conventional decision algorithms like model predictive control (MPC)~\cite{kouvaritakis2016model} computationally infeasible. 

% methods
To address these challenges, we propose a system composed of three key modules. First, we develop a \textbf{gradient-based prompt inversion module} (\S \ref{sec:method-prompt-inv}) that converts video frames into compact semantic representations. By optimizing text embeddings through backpropagation, we generate a small set of learned tokens that, \emph{together with a short textual prompt}, can be transmitted to the client and decoded into high-quality frames using stable diffusion. Second, we design a \textbf{computation-aware scheduling strategy} (\S \ref{sec:score-strategy}) that accounts for both bandwidth and decoding latency. We extend the definition of stall time to include decoding delays, and introduce an out of order downloading mechanism that improves hardware utilization by overlapping prompt decoding, H.265 decoding, and network transfer across different compute units. Finally, we introduce a \textbf{tree based search algorithm} (\S \ref{sec:large-decision-space-searching}) that reduces planning overhead. By combining Monte Carlo Tree Search with pruning techniques, we efficiently explore the exponentially large decision space introduced by multiple codecs, bitrate, decoding latencies, and download orders.

% results
The results are encouraging. With all three modules working together, our system significantly improves the performance of short video preloading. Compared to existing strategies, it reduces stalls and bandwidth waste by over 31\%, while improving average user QoE by 45\%. These gains demonstrate the practical potential of using local computation to improve the efficiency and quality of short video preloading.

\vspace{-2mm}
\section{Motivation and Challenges}
\vspace{-1mm}

\subsection{Stalls and Waste Are Still Inevitable}

Video preloading is widely used in short video platforms to improve playback experience~\cite{phong2023joint,zhang2020apl}. By predicting what the user is likely to watch next, and downloading part of the video in advance, the system aims to reduce stalls and ensure smooth viewing. However, in practice, preloading still suffers from two problems: playback stalls and bandwidth waste~\cite{phong2023joint}. 

The root cause of the stall–waste dilemma lies in prediction errors: it is difficult to anticipate how long a user will continue watching a video~\cite{zhang2023user} or how much bandwidth will be available in the future~\cite{kurdoglu2016real}. As a result, preloading decisions often involve a binary tradeoff. Preloading too little risks playback stalls when future chunks are not delivered in time~\cite{le2013buffer}, while preloading too much wastes bandwidth if the user scrolls away early~\cite{zhang2022measurement}. The system must choose between competing objectives, often sacrificing one to protect the other. 

Previous approaches~\cite{phong2023joint,zhang2020apl} have tried to improve preloading decisions by designing better predictors or scoring rules. But as long as the decision space remains binary, this tradeoff cannot be fundamentally resolved. 

\vspace{-1mm}

\subsection{Can We Reduce Both Stalls and Waste?}
\vspace{-1mm}

But what if we could reduce that cost without lowering quality? Our key insight is that reducing chunk sizes without compromising perceptual quality can simultaneously lower both playback stalls and bandwidth waste. 

To explore this idea, we conducted an experiment on the PDAS~\cite{zhou2022pdas} simulator, a widely used evaluation framework for short video preloading systems. The only change we made was to reduce the size of each video chunk in the dataset, while keeping its quality metrics, the user swipe traces, the bandwidth traces, and the preloading algorithm unchanged. This simulates a system where videos require less bandwidth without sacrificing visual quality. 

% \vspace{-4mm}
\begin{figure}[t]
    \centering
    \includegraphics[width=0.8\linewidth]{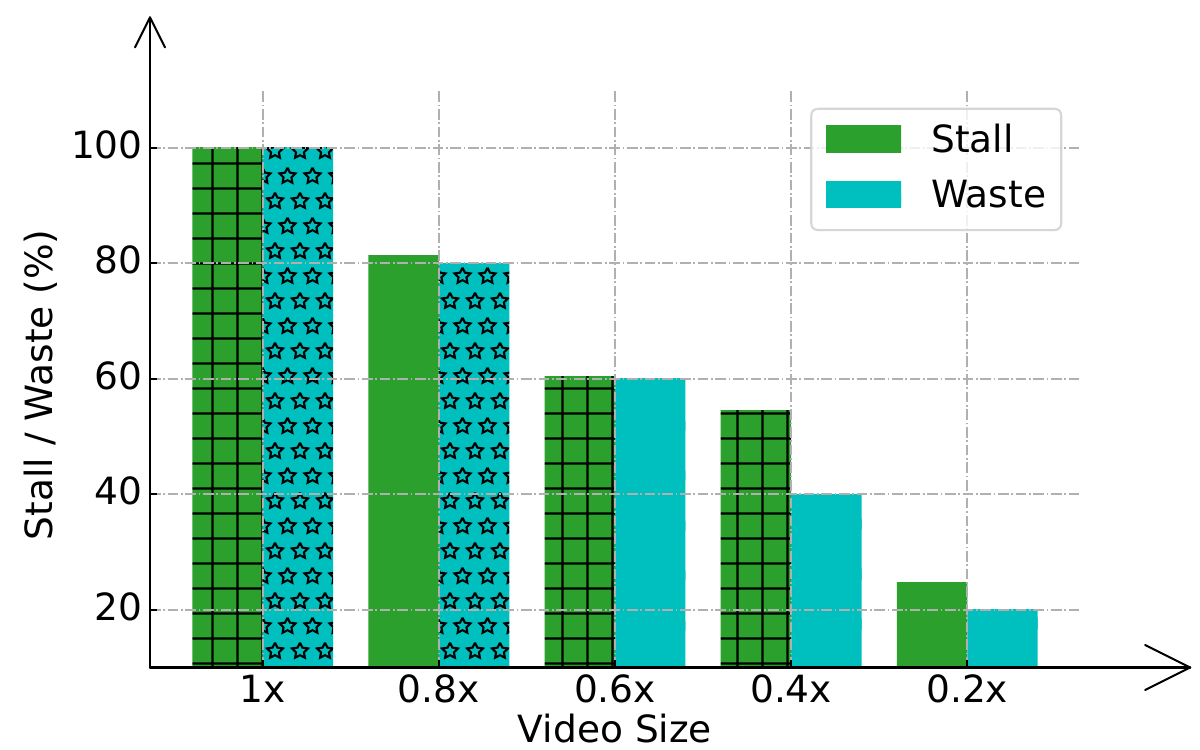}
    \vspace{-2mm}
    \caption{
        Impact of data budget on stall and bandwidth waste.
        % With smaller video sizes, the system preloads more flexibly, resulting in significantly fewer stalls and less wasted data. 
    }
    \vspace{-5mm}
    \label{fig:stall-waste}
\end{figure}

The results were striking. As shown in Fig. \ref{fig:stall-waste}, smaller chunk sizes can lead to over 70\% reduction in both stall time and bandwidth waste. In addition, since each chunk consumed less bandwidth, the system had more headroom to select a higher bitrate without exceeding the bandwidth budget. 

This suggests that if we can find a way to shrink chunk sizes without hurting perceptual quality, we open up a new opportunity: improving both performance and user experience. 
% If chunks are smaller, the system can preload more content within the same bandwidth budget, reducing the risk of stalls. At the same time, when users leave early, the amount of unused data is much smaller, minimizing waste. 

\begin{figure*}[t]
  \centering
  \includegraphics[width=\textwidth]{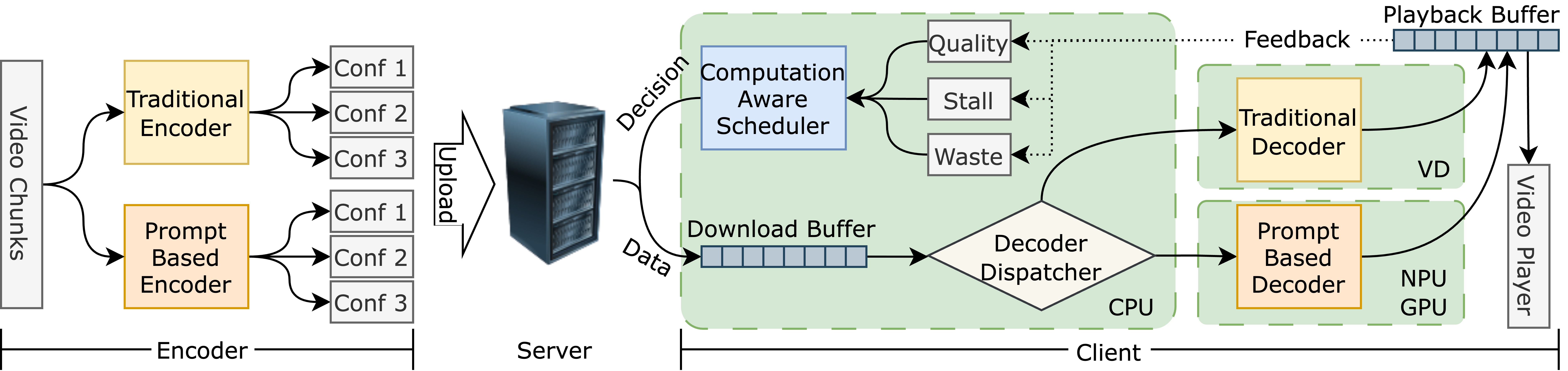}
  \caption{
    \textbf{System Overview.} Our system extends traditional short video preloading by introducing prompt-based encoding and computation-aware scheduling. The system includes three roles: an \emph{encoder} that generates both conventional (H.265) and prompt-based representations; a \emph{server} that stores multiple options per chunk; and a \emph{viewer} that uses a computation-aware scheduler to decide which chunks to download and decode. The viewer exploits parallelism across CPU, GPU, NPU, and Video Decoder(VD), with a Decoder Dispatcher routing each chunk to the appropriate backend.
  }
  \label{fig:system-overview}
  \vspace{-4mm}
\end{figure*}

\subsection{Use Computation to Break the Tradeoff}

But how? While super-resolution\cite{yeo2018neural} or more advanced neural codecs\cite{jia2025towards} can also trade local computation for improved quality, we take this idea a step further. Instead of transmitting pixel-level encodings, we send compact \emph{semantic prompts} that can be decoded by modern generative models~\cite{Rombach_2022_CVPR} to reconstruct high-quality frames, offering greater potential for quality enhancement under the same bandwidth constraints. More importantly, many of them are fully open source and can be optimized to run on modern mobile hardware, including GPUs and NPUs already present in consumer smartphones~\cite{stable-diffusion-coreml-apple-silicon}. 
% So instead of transmitting traditional video compressed bitstreams, we transmit a compact semantic representation, such as a prompt, and let a generative model on the device reconstruct the visual content. 

By introducing local computation that is already available on most user devices into the preloading process, we can now trade computation for reduced transmission, instead of relying solely on bandwidth and prediction accuracy. 
% This creates the opportunity to improve both playback smoothness and bandwidth waste, using a new resource that is already available on most user devices.

\vspace{-2mm}
\subsection{Challenges}

While the idea of using computation to reduce bandwidth is appealing, realizing it in a practical short video preloading system presents several significant challenges. 
% Below, we outline the key technical problems and provide an overview of how our system addresses them. 

% \vspace{-1mm}

\noindent \paragraph{Challenge 1: Constructing Compact and Effective Semantic Representations.}  
Replacing traditional video data with semantic inputs requires a way to encode each frame into a minimal representation that still supports high-fidelity reconstruction. We address this by designing a gradient based prompt inversion module, described in Section \ref{sec:method-prompt-inv}. 

% \vspace{-1mm}

\paragraph{Challenge 2: Scheduling Under Computation and Bandwidth Constraints.}  
Unlike conventional codecs, decoding a prompt-based representation with a diffusion model introduces significant latency, fundamentally changes how preloading decisions should be made. When decoding becomes a bottleneck, the scoring function must also incorporate compute cost. We address the challenge by developing a computation-aware scheduler that reorders downloads and adapts chunk prioritization based on both bandwidth and compute availability, as described in Section \ref{sec:score-strategy}.

% \vspace{-1mm}

\paragraph{Challenge 3: Searching Over a Large and Flexible Decision Space.}  
Once we introduce computation into the preloading pipeline, we must reason over a time window to understand how a sequence of download and decode decisions will interact with each other. Moreover, our new chunk scoring mechanism brings additional complexity due to prompt-augmented options per bitrate and can be downloaded and decoded out of order. To tackle this, we design a structured decision search procedure based on Monte Carlo tree search, augmented with pruning strategies that discard unpromising branches early, as described in Section \ref{sec:large-decision-space-searching}.

\vspace{-2mm}

\section{System Overview}
\vspace{-1mm}

As shown in Figure~\ref{fig:system-overview}, our system extends the traditional video transmission pipeline by introducing two new components: a prompt-based encoder-decoder pair and a computation aware decision module. These additions allow the system to leverage idle compute resources to optimize the preloading performance. The system is organized around three roles: an \emph{encoder} that prepares multiple representations of each video chunk, including both conventional H.265 and prompt based encodings; a \emph{server} that stores all candidate versions along with their metadata; and a \emph{client} that decides what to download, in which order, and when to decode it based on network and compute conditions. The following describes the encoder and viewer pipelines in detail. 

% \vspace{-2mm}

\paragraph{Encoding Pipeline}
Given a video, the encoder generates multiple versions of each chunk. The first is a set of conventional H.265 bitstreams at various bitrates, produced by a standard video encoder. The second is a prompt based representation obtained through our gradient based prompt inversion module, which produces compact text embeddings capable of reconstructing the video frames via a diffusion model. All representations are uploaded to the server, where they are stored along with metadata such as decoding latency, bitrate, and estimated quality. 

% \vspace{-3mm}

\paragraph{Client Scheduler and Decoder}
When watching videos, the client is responsible for deciding which chunks to download, in which order, and in which format. These decisions are made by our computation aware scheduler, which considers network bandwidth, compute capacity, decoding latency, and playback deadlines. Chunks encoded with H.265 are decoded using the CPU and dedicated Video Decoding(VD) units, while prompt-based chunks are reconstructed on the GPU or NPU via the diffusion model. Since these hardware resources operate independently, the system exploits this parallelism to improve utilization and reduce stalls. The scheduler is free to download chunks out of order, and may pre-decode prompt-based keyframes in parallel with downloading traditional ones, enabling flexible playback planning.

% \section{Prompt Inversion: What and How}
\vspace{-2mm}
\section{Prompt Inversion}
\label{sec:method-prompt-inv}
\vspace{-2mm}

In this section, we detail our approach to generating compact semantic representations of video frames using a gradient-based prompt inversion framework built upon stable diffusion. We then explain our design choices regarding which components to invert for minimal bandwidth usage and high perceptual quality. Finally, we describe how we integrate prompt encoding with conventional H.265 compression to accelerate decoding and reduce computational overhead.

\vspace{-2mm}

% \subsection{How? Gradient based Prompt Inversion}
\subsection{Gradient based Prompt Inversion}
\vspace{-1mm}

To enable semantic video compression, we leverage the stable diffusion generative model, which synthesizes images by denoising random noise under a textual condition. This textual condition is constructed by embedding tokens from a vocabulary into word vectors by using a tokenizer~\cite{wolf2020transformers} module. Those word embeddings are then processed by a CLIP~\cite{radford2021learning} model to produce a sentence level embedding aligned with the image space. The pipeline including the CLIP and the diffusion denoising module is fully differentiable. 

\begin{figure}[t]
  \centering
  \includegraphics[width=0.45\textwidth]{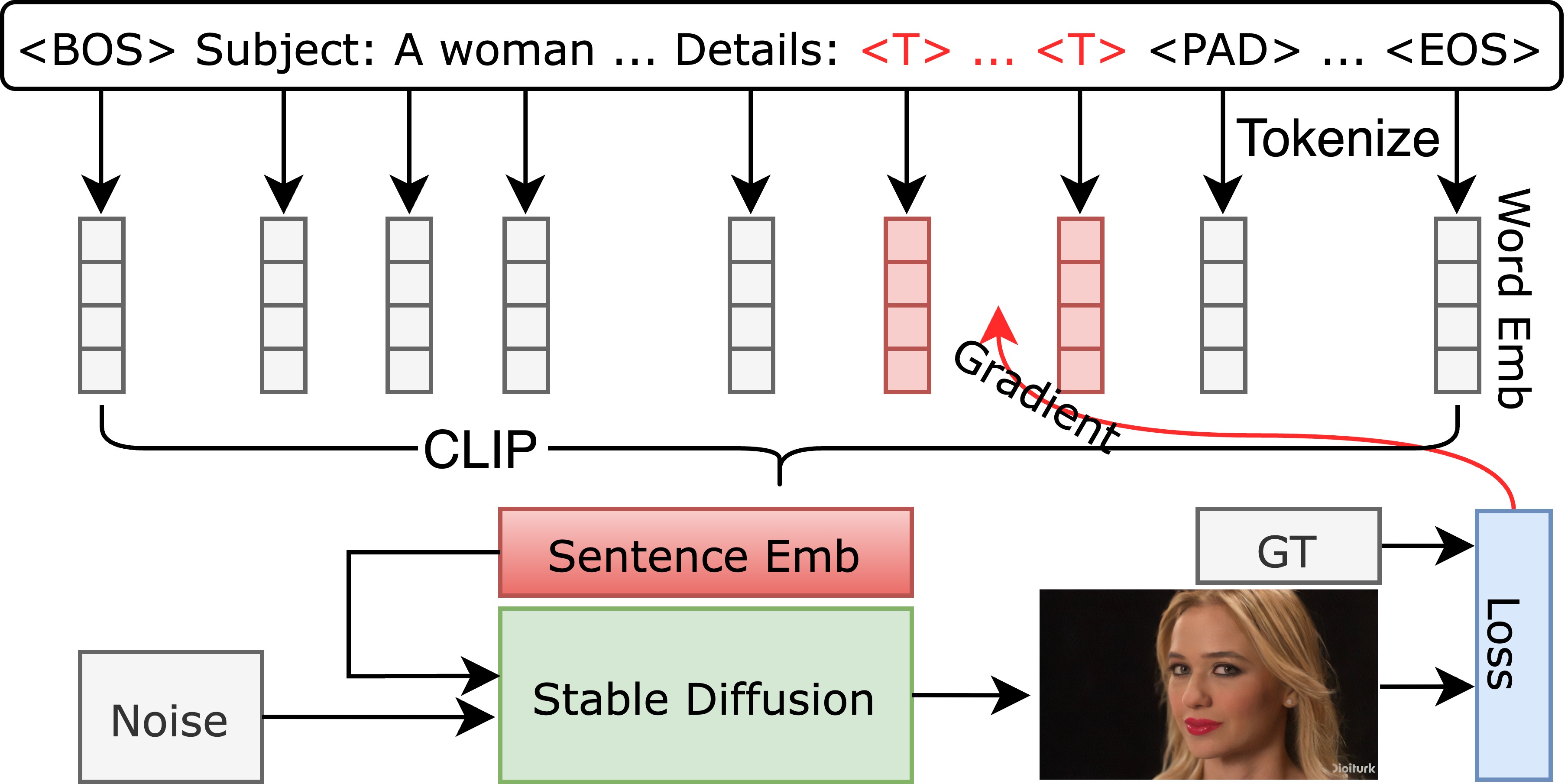}
  \vspace{-2mm}
  \caption{Overview of the encoding pipeline.}
  \label{fig:method-sd-pipe}
  \vspace{-7mm}
\end{figure}

As illustrated in Figure~\ref{fig:method-sd-pipe}, our encoding pipeline begins by generating a textual description of the input frame using an image-to-text module~\cite{li2022blip}. We then construct a full stable diffusion prompt by appending some learnable tokens, denoted as \texttt{<T>}, to this description. The embeddings of these learnable tokens are the target of our optimization process.

This prompt is passed through the standard stable diffusion pipeline. It is first tokenized via the \emph{Tokenizer}, then encoded by CLIP module into a sentence level representation. This sentence embedding serves as the textual condition for the diffusion model, which denoises a fixed random noise vector to generate a reconstructed image. Using a fixed noise seed ensures that the gradients remain stable and consistent across iterations. The reconstructed image is then compared against the original input frame using both pixel-wise MSE loss and perceptual LPIPS loss. Gradients are backpropagated through the diffusion model and the CLIP encoder, but only the embeddings of the learnable \texttt{<T>} tokens are updated. After sufficient iterations, these tokens can reconstruct a high quality approximation of the original frame.

\vspace{-2mm}
\subsection{What to Invert? Sentence Or Token?}
% \vspace{-1mm}

% \vspace{4mm}
\begin{figure}[t]
\vspace{4mm}
  \centering
  \includegraphics[width=0.33\textwidth]{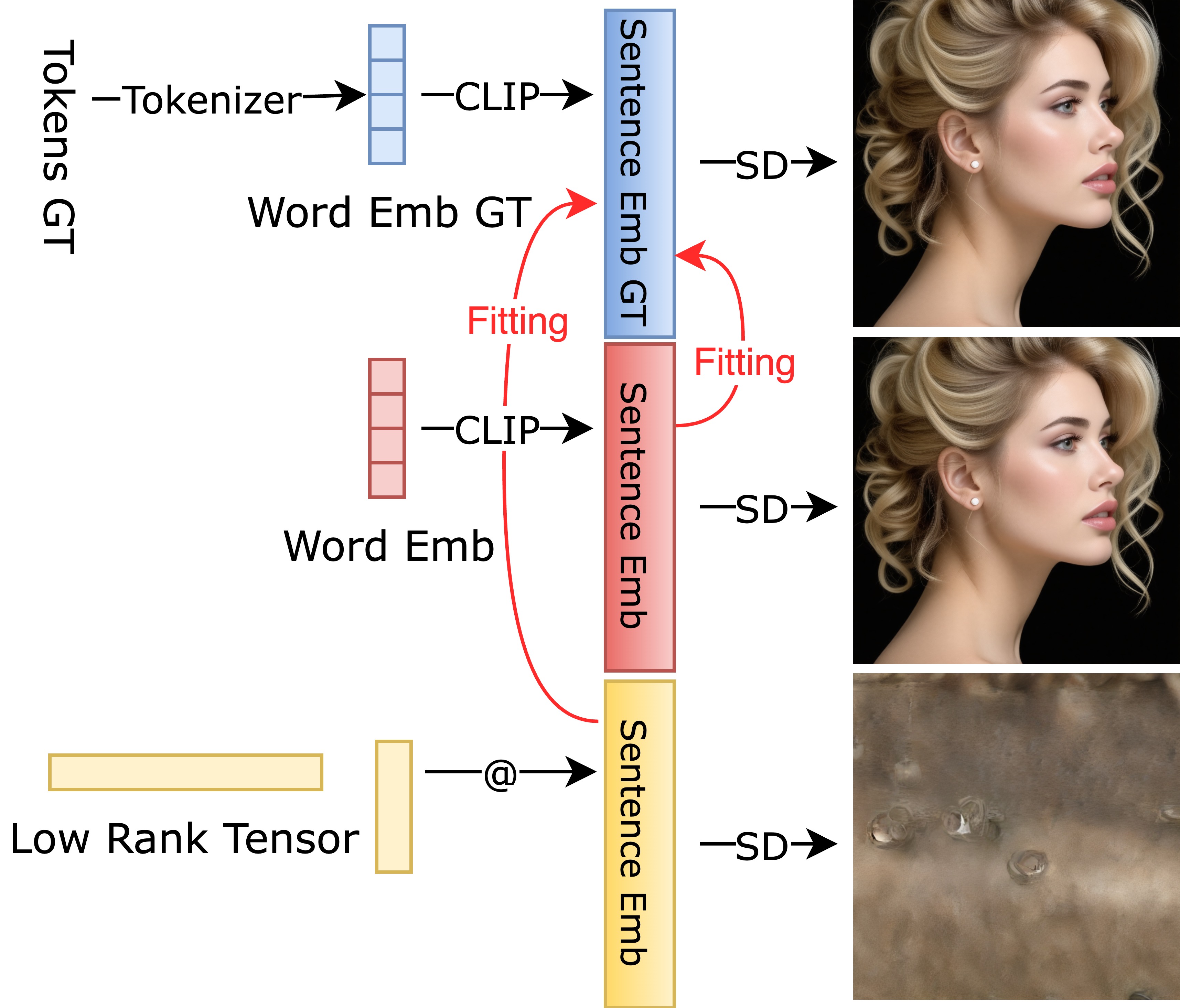}
  \caption{Reconstructions from low rank approximations of sentence embeddings fail to preserve fine details, indicating that the low rank assumption does not hold.}
  \vspace{-3mm}
  \label{fig:method-ass-not-hold}
  \vspace{-4mm}
\end{figure}

A critical question is which level of textual condition to invert for best compression and reconstruction quality. Prior work~\cite{wu2024promptus} suggested that we can inverting the sentence-level embedding, assuming it has a low-rank structure that allows dimensionality reduction via matrix factorization. 
% They further reduced bitrate by only transmitting prompts for keyframes and interpolating others.

We conducted experiments that challenge this low rank assumption. First, as shown in Figure. \ref{fig:method-ass-not-hold}, we found that attempting to approximate a sentence embedding with a low-rank tensor produces reconstructions that are abstract and diverge significantly from the original image, indicating the embedding lacks a low-rank structure. Second, we demonstrated that a small set of token-level embeddings, only four tokens with much less data than a full sentence embedding, can be optimized to reconstruct the image nearly as well as the original sentence embedding. These findings suggest that inverting token-level embeddings rather than sentence embeddings yields better compression and higher reconstruction fidelity, as token embeddings better conform to the distribution expected by the diffusion model.

An additional advantage of optimizing word-level embeddings is that they are in the same space as textual prompts. This allows the system to leverage the strong semantic priors and compositional capabilities of both CLIP and the underlying language model, enabling more expressive and interpretable reconstructions. By jointly transmitting a short text prompt and a small set of optimized token embeddings, we can achieve higher reconstruction quality with less data.

\vspace{-2mm}

\subsection{Speed Up Decoding by Hybrid Coding}
\vspace{-1mm}

While using prompt-based inversion for every frame can achieve high compression, it is computationally intensive and thus impractical for real-time decoding on resource constrained devices such as mobile phones~\cite{liu2025promptmobile}. However, as shown in Table \ref{tab:eval-iframe-quality-size}, we observe that when matched for visual quality (e.g., LPIPS~\cite{johnson2016perceptual}), prompt embeddings for keyframes occupy significantly fewer bits compared to traditional I-frames encoded with H.265~\cite{265}. This insight motivates a hybrid strategy: we encode only the I-frames using prompt-based inversion, and use conventional low-bitrate B/P-frame compression for the rest. 
% \vspace{-4mm}

% \vspace{5mm}
\begin{table}[htbp]
\vspace{4mm}
\caption{Comparison of I-frame quality (LPIPS ↓) and frame size (KB) under different target bitrates. Our keyframes consistently achieve better quality-to-size tradeoffs.}
% compared to H.265-generated I-frames.}
\label{tab:eval-iframe-quality-size}
\centering
\vspace{-1mm}
\begin{tabular}{lcccc}
\toprule
\textbf{\makecell{Bitrate\\(kbps)}} & \multicolumn{2}{c}{\textbf{Ours}} & \multicolumn{2}{c}{\textbf{H.265 I-frame}} \\
\cmidrule(r){2-3} \cmidrule(l){4-5}
 & LPIPS ↓ & Size (KB) & LPIPS ↓ & Size (KB) \\
\midrule
200   & 0.459 & 8.8 & 0.543 & 7.4 \\
400   & 0.459 & 8.8 & 0.510 & 11.2 \\
600   & 0.459 & 8.8 & 0.494 & 14.8 \\
1200  & 0.459 & 8.8 & 0.454 & 27.7 \\
\bottomrule
\end{tabular}
\vspace{-6mm}
\end{table}
% \vspace{-4mm}

% \vspace{-3mm}

Our experiments show that this mixed encoding scheme achieves over 50\% reduction in bitrate compared to pure H.265 encoding at the same perceptual quality, providing a practical tradeoff between computational cost and bandwidth efficiency.

\vspace{-2mm}
\section{Computation Aware Strategy}
\label{sec:score-strategy}

\subsection{Scoring for Chunks}

Prompt-based decoding offers excellent compression but comes with substantial latency. Decoding a 1s 1080p chunk can take over 1400ms on mobile GPUs or NPUs, compared to under 1ms for H.265 decoding. This prohibits a simple replacement of all chunks with prompt-based formats. Instead, the system must carefully mix prompt and H.265 chunks and schedule them to meet playback deadlines. 

So we define a computation aware scoring function that jointly consider bandwidth, computation, and playback deadlines to guide preloading decisions. Each chunk $i$ is associated with four metrics: visual quality $q_i$, quality variation $v_i = |q_i - q_{i-1}|$, stall duration $\sigma_i$, and bandwidth cost $b_i$. These metrics capture user perceived fidelity, smoothness, and bandwidth consumption.
Stall duration $\sigma_i$ is computed based on download, decode, and playout scheduling. Let $\delta_i$ be the download duration, $\chi_i$ the decode time, and $\pi_i$ the playout duration for chunk $i$. Let $\Delta_i$, $\Gamma_i$, $\beta_i$, and $\Phi_i$ denote the download end time, decode end time, buffer availability time, and playback start time, respectively. These are computed recursively as:
\vspace{-3mm}

% \begin{equation}
% \[
% \Delta_i = \Delta_{i-1} + \delta_i, \quad \Delta_0 \triangleq 0
% \]
% \[
% \Gamma_i = \max(\Delta_i,\; \Gamma_{i-1}) + \chi_i, \quad \Gamma_0 \triangleq 0
% \]
% \[
% \beta_i = \Phi_{i-1} + \pi_{i-1}, \quad \beta_1 \triangleq 0
% \]
% \[
% \Phi_i = \max(\Gamma_i,\; \beta_i)
% \]
% \end{equation}

% \vspace{-3mm}
\begin{equation}
\Delta_i = \Delta_{i-1} + \delta_i, \quad \Delta_0 \triangleq 0
\vspace{-6mm}
\end{equation}

\begin{equation}
\Gamma_i = \max(\Delta_i,\; \Gamma_{i-1}) + \chi_i, \quad \Gamma_0 \triangleq 0
\vspace{-2mm}
\end{equation}

We then define stall duration as the excess delay beyond buffer availability:

\vspace{-4mm}
\begin{equation}
\sigma_i = \max(0, \Gamma_i - \beta_i)
\vspace{-2mm}
\end{equation}

Using these definitions, we compute the chunk-level score $f_i$ as a weighted sum:

\vspace{-4mm}
\begin{equation}
f_i = w_1 q_i - w_2 v_i - w_3 \sigma_i - w_4 b_i
\vspace{-2mm}
\end{equation}
where $w_1$–$w_4$ are tunable weights reflecting system preferences among quality, smoothness, latency, and efficiency.

Finally, our goal is to select a scheduling plan that maximizes the total utility across all video chunks. Formally, we search for the optimal plan \( \mathcal{P}^* \) such that:

\vspace{-3mm}
\begin{equation}
\mathcal{P}^* = \arg\max_{\mathcal{P}} \sum_{i=1}^{N} f_i
\vspace{-1mm}
\end{equation}

\vspace{-2mm}
\subsection{Out of Order, Outperforms}
\vspace{-1mm}

\begin{figure}[t]
  \centering
  \begin{subfigure}{0.75\linewidth}
    \centering
    \includegraphics[width=\linewidth]{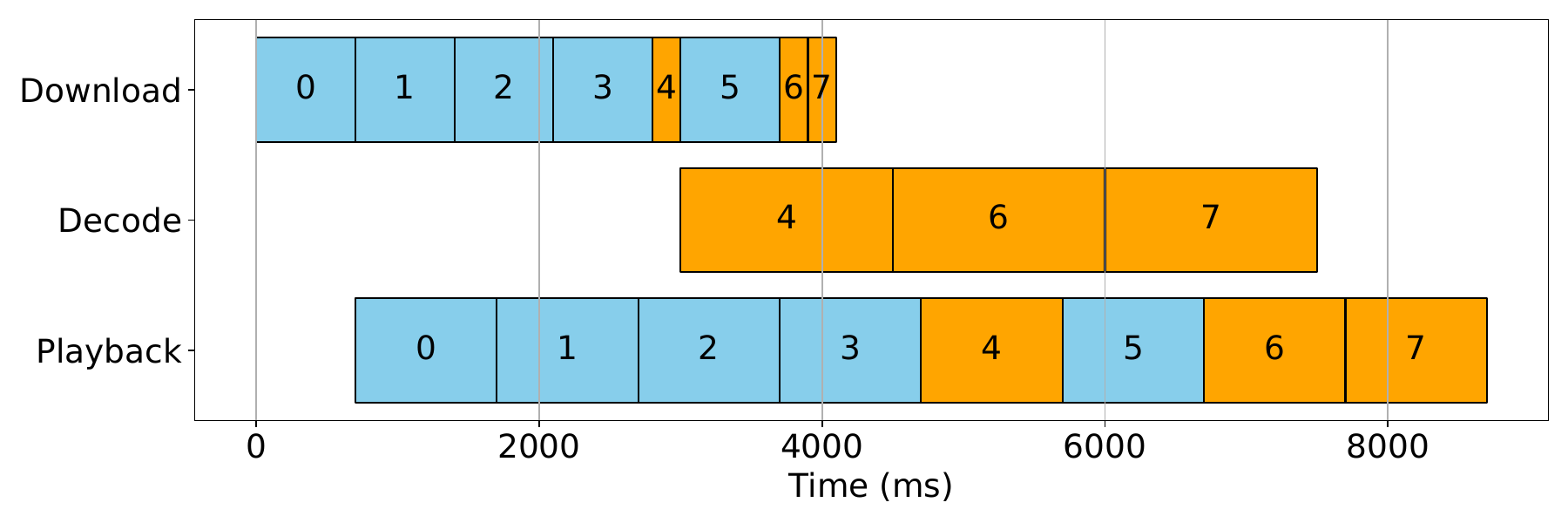}
    \vspace{-7mm}
    \caption*{\textbf{(a)} Sequential Scheduling}
  \end{subfigure}
  \vspace{0.5em}
  \begin{subfigure}{0.75\linewidth}
    \centering
    \includegraphics[width=\linewidth]{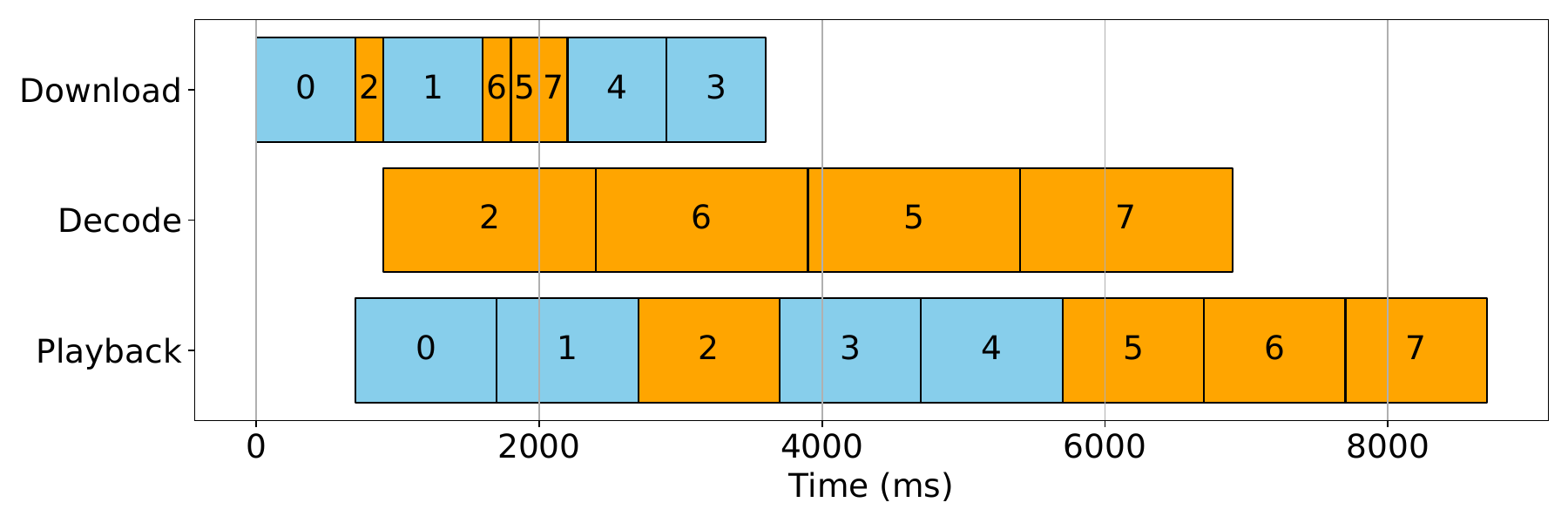}
    \vspace{-7mm}
    \caption*{\textbf{(b)} Out of Order Scheduling}
  \end{subfigure}
  \vspace{-3mm}
  \caption{
    Comparison of sequential and out of order chunk loading strategies. Out of order scheduling enables earlier and more efficient prompt chunk decoding, allowing more content to be served in the prompt format, reducing bandwidth usage and improving compute utilization.
    }
  \label{fig:timeline-comparison}
  \vspace{-5mm}
\end{figure}

However, once large decoding delays are introduced, traditional sequential downloading strategies become inefficient. Our key insight is that modern mobile hardware supports multiple concurrent compute units: CPU and video decoders handle H.265 decoding, while GPU/NPU handle prompt decoding, and downloading is managed by the CPU. Thus, downloading and decoding can proceed in parallel, making scheduling chunks more efficiently.

Consider a toy example where each video chunk represents 1000ms of playback. A prompt-based chunk requires 200ms to download and 1500ms to decode, while a H.265 chunk takes 700ms to download and has negligible decode latency. We compare two scheduling strategies: one that downloads chunks strictly in playback order, and another that allows out of order selection to maximize compute and bandwidth utilization. 

As shown in Figure. \ref{fig:timeline-comparison}, under the sequential strategy, the system downloads 5 H.265 chunks and only 3 prompt chunks within the time window. Prompt decoding cannot begin until playback buffer has been long enough, leading to under utilized GPU/NPU resources and delayed prompt usage. In contrast, the out of order scheduler starts downloading and decoding prompt chunks earlier. This enables the system to decode 4 prompt chunks in the same time budget, effectively shifting more content to the compact format. 

% Beyond timeline efficiency, our out of order scheduler also delivers significant end-to-end improvements across key metrics. As shown in Figure~\ref{fig:compare_ratio}, compared to the PDAS baseline, our method reduces stall time, wasted data, and overall video size, while also improving inverse QoE. These gains reflect better utilization of both bandwidth and on-device compute, and demonstrate the effectiveness of jointly optimizing decoding cost and scheduling order.

% \section{Planning Over a Large Decision Space with MCTS and Pruning}
\vspace{-2mm}
\section{Planning with MCTS and Pruning}
\label{sec:large-decision-space-searching}
\vspace{-2mm}

% To address the challenge of searching over a large and flexible decision space, we design a planning algorithm based on Monte Carlo Tree Search (MCTS). Unlike conventional preloading systems that select the next chunk greedily, our system must plan ahead across a horizon of multiple chunks, evaluating the joint impact of download and decoding configurations on future playback. This is particularly critical when prompt decoding latency is high and chunks can be downloaded and decoded out of order.

Our system must make preloading decisions over a planning horizon of multiple future chunks. For each chunk $c_i$, there is a set of candidate configurations $R_i = \{r^i_1, r^i_2, \dots\}$, corresponding to different encodings (e.g., H.265 vs. prompt), bitrates, and decoding latencies. The total number of possible plans grows exponentially as $\prod_{i=1}^H |R_i|$, making exhaustive enumeration infeasible. Traditional model predictive control (MPC) methods, which explore a limited number of candidate trajectories, are insufficient for this scale. 

To tackle this challenge, we design a planning algorithm based on Monte Carlo Tree Search (MCTS)~\cite{browne2012survey}. We define the planning horizon as $H$ chunks. At each step, the planner samples promising decision paths using the UCT rule~\cite{kocsis2006uct}: 
\vspace{-3mm}

\begin{equation}
\text{UCT}(s) = \frac{V(s)}{n(s)} + \alpha \cdot \sqrt{\frac{\log N}{n(s)}}
\vspace{-2mm}
\end{equation}

where $n(s)$ is the number of times $s$ has been visited, $N$ is the total number of simulations from the root, $\alpha$ is a tunable exploration parameter, and its reward $V(s)$ is computed as the sum of chunk scores from the root to that node $s$:

% \vspace{-2mm}
\begin{equation}
% \vspace{-2mm}
V(s) = \sum_{j=1}^{i} f_j
\vspace{-2mm}
\end{equation}

To further reduce the search space, we apply two key techniques: aggressive pruning and early stopping. 

First, during node expansion, we prune any configuration that causes playback stalls due to delayed decoding. Specifically, a child node is discarded if:

\vspace{-4mm}
\begin{equation}
\text{ComputeStall}_j > 0 \quad \text{for any } j \leq i
\vspace{-2mm}
\end{equation}
This strict pruning guarantees that no partial plan in the tree introduces stalls from computation delays, ensuring only feasible paths are explored.

Second, we introduce early stopping to ensure timely decision making under runtime constraints. If the remaining time in the planning cycle is insufficient for further exploration, the planner can immediately select the best complete or partial plan observed so far. This allows the system to produce high quality schedules within tight runtime budgets.

\begin{figure}[t]
    \centering
    \includegraphics[width=0.65\linewidth]{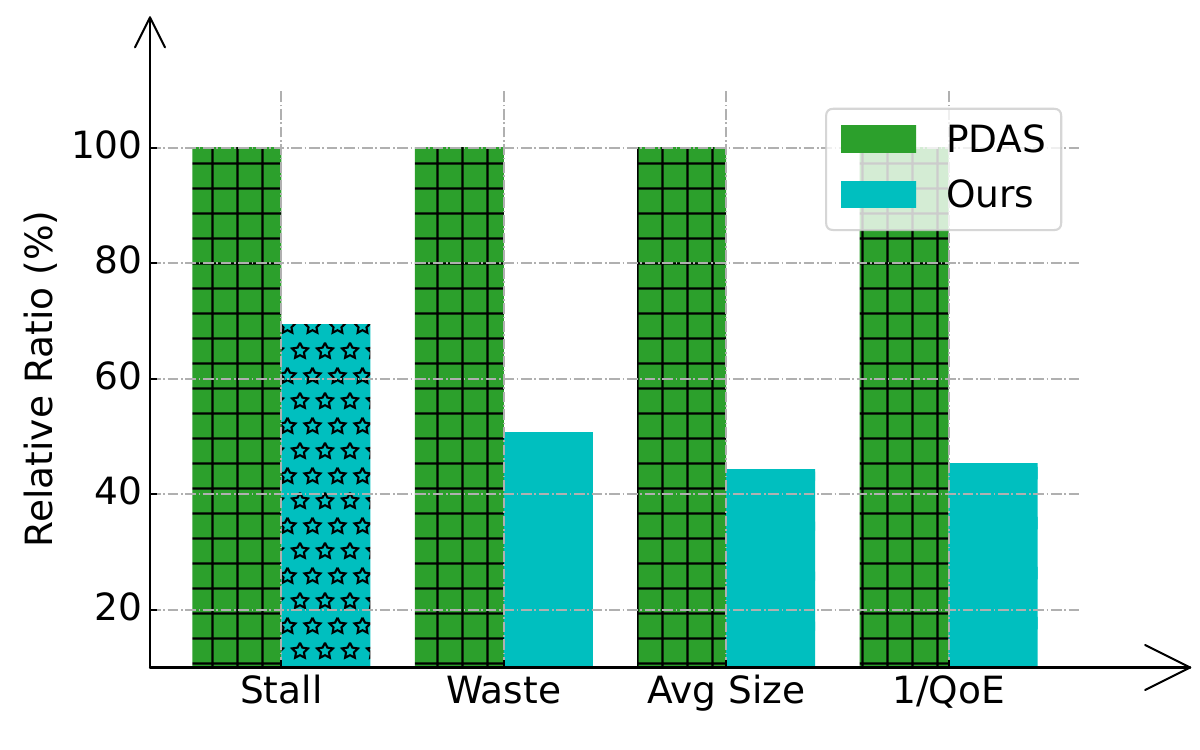}
    \vspace{-2mm}
    \caption{Comparison of normalized metrics between PDAS and our approach. All PDAS values are normalized to 100\%. Our method achieves significantly lower stall, waste, video size and inverse QoE values, indicating better overall performance.}
    \vspace{-10mm}
    \label{fig:compare_ratio}
\end{figure}

\vspace{-2mm}
\section{Performance Evaluation}
\vspace{-3mm}

% \subsection{Setup}

\subsection{Video Quality}
\label{sec:video_quality}
\vspace{-2mm}

\begin{figure}[t]
    \centering
    \includegraphics[width=0.63\linewidth]{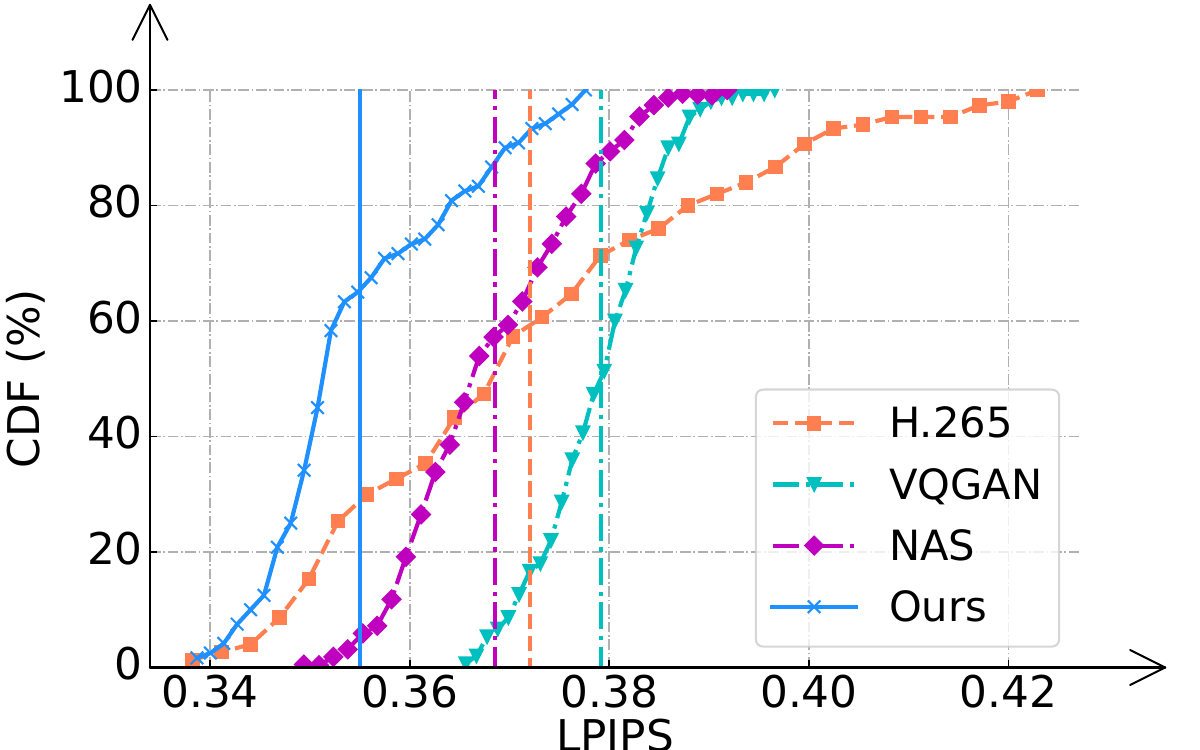}
    \vspace{-2mm}
    \caption{CDF of LPIPS scores under an average bitrate of 280~kbps. Our method reduces the proportion of low-quality frames by 40\% compared to H.265.}
    \label{fig:lpips_cdf}
    \vspace{-5mm}
\end{figure}

We compare our approach with three codecs: the traditional H.265~\cite{265}, the neural codec VQGAN~\cite{esser2021taming}, and the super-resolution–based codec NAS~\cite{yeo2018neural}.  
All methods are evaluated under the same average bitrate constraint of 280~kbps.  
As shown in the cumulative distribution function (CDF) of LPIPS~\cite{zhang2018unreasonable} scores in Figure~\ref{fig:lpips_cdf}, our method significantly improves perceptual quality.  
Specifically, the proportion of frames with LPIPS greater than 0.37 is reduced by 40\% compared to H.265, indicating a substantial improvement in visual fidelity. 
% under the same bandwidth budget.

\vspace{-4mm}
\subsection{Preloadng Performance}
\vspace{-2mm}

We use the trace provided by the Grand Challenge\cite{MMGC2022} to evaluate our scheduler, which delivers significant end-to-end improvements across key metrics. As shown in Figure~\ref{fig:compare_ratio}, compared to the PDAS baseline, our method reduces stall time, wasted data, and overall video size, while also improving QoE. These gains reflect better utilization of both bandwidth and on-device compute, and demonstrate the effectiveness of jointly optimizing decoding cost and scheduling order.

\vspace{-2mm}
\subsection{Module Ablation}
\vspace{-1mm}

\noindent \textbf{Word-level Embedding Optimization.}
This refinement leads to improved perceptual quality: the average LPIPS decreases from 0.372 (sentence-level) to 0.356 (word-level), confirming that more precise semantic alignment effectively enhances frame reconstruction fidelity.

\noindent \textbf{Monte Carlo Tree Search.}  
When the prediction window is set to 7, the brute-force enumeration requires exploring all \(7! \times 6^7 \approx 1.4\times10^9\) possible order–configuration combinations.  
In contrast, our Monte Carlo Tree approach converges to the same optimal solution within only \(1.2\times10^5\) evaluations, achieving a speedup of over \(10^4\times\).  
This reduction in search steps directly translates to lower computational cost. 

\vspace{-2mm}
\section{Conclusion}
\vspace{-1mm}

This paper presents a new paradigm for short video preloading that leverages local computation to reduce bandwidth requirements and mitigate the fundamental stall–waste dilemma. By exploiting the largely idle compute resources on user devices, our system transmits compact semantic prompts instead of pixel-level encodings, allowing generative models to reconstruct high-quality frames under strict bitrate constraints. We design three key modules to make this vision practical: a gradient-based prompt inversion module for semantic compression, a computation-aware scheduler for coordinated bandwidth–latency optimization, and a tree-based search algorithm that efficiently explores the vast decision space of codecs and decode orders. Experimental results show that our approach reduces stalls and bandwidth waste by over 31\% and improves user QoE by 45\% compared to existing preloading strategies. We believe this work opens a promising direction toward computation-assisted video delivery. This work was sponsored by the NSFC grant(62431017), Bytedance Grant(CT20241126107484). We gratefully acknowledge the support of State Key Laboratory of Media Convergence Production Technology and Systems, Key Laboratory of Intelligent Press Media Technology. The corresponding author is Xinggong Zhang(zhangxg@pku.edu.cn). 

\vspace{-2mm}

\bibliographystyle{IEEEtran} 
\bibliography{hotnets25-template}

\end{document}